# Direct assessment of the sensitivity drift of SQM sensors installed outdoors

Salvador Bará[1,*], Enric Marco[2], Salvador J. Ribas[3], Manuel Garcia Gil[4], Alejandro Sánchez de Miguel[5], Jaime Zamorano[6]

[1] Departamento de Física Aplicada, Universidade de Santiago de Compostela, 15782 Santiago de Compostela, Galicia
[2] Departament d'Astronomia i Astrofísica, Facultat de Física, Universitat de València, 46100 Burjassot, Spain
[3] Parc Astronòmic Montsec - Ferrocarrils de la Generalitat de Catalunya. Camí del Coll d'Ares, s/n, Àger (Lleida) 25691, Spain, and Departament de Física Quàntica i Astrofìsica. Institut de Ciències del Cosmos (ICC-UB-IEEC), C Martí Franquès 1, Barcelona 08028
[4] Departament de Territori i Sostenibilitat, Generalitat de Catalunya, and Departament de Projectes d'Enginyeria, Universitat Politècnica de Catalunya / BARCELONATECH, Spain
[5] Environment and Sustainability Institute, University of Exeter, Penryn, Cornwall, TR10 9FE, United Kingdom, and Instituto de Astrofísica de Andalucía, Glorieta de la Astronomía, s/n, C.P. 18008, Granada, Spain.
[6] Departmento Física de la Tierra y Astrofísica and Instituto de Física de partículas y del Cosmos IPARCOS, Universidad Complutense de Madrid, Madrid, 28040, Spain.

**Abstract**

Long-term monitoring of the evolution of the artificial night sky brightness is a key tool for developing science-informed public policies and assessing the efficacy of light pollution mitigation measures. Detecting the underlying artificial brightness trend is a challenging task, since the typical night sky brightness signal shows a large variability with characteristic time scales ranging from seconds to years. In order to effectively isolate the weak signature of the effect of interest, determining the potential long term drifts of the radiance sensing systems is crucial. If these drifts can be adequately characterized, the raw measurements could be easily corrected for them and transformed to a consistent scale. In this short note we report on the progressive darkening of the signal recorded by Sky Quality Meter (SQM) detectors belonging to several monitoring networks, permanently installed outdoors for periods ranging from several months to several years. The sensitivity drifts were estimated by means of parallel measurements made at the beginning and at the end of the evaluation periods using reference detectors of the same kind that were exposed to little or no weathering in the intervening time. Our preliminary results suggest that SQM detectors installed outdoors steadily increase (darken) their readings at an average rate of +0.036 $mag_{SQM}/arcsec^2$ per $MWh/m^2$ of exposure to solar horizontal global irradiation, that for our locations at mid latitudes (39º-43º N) in temperate climate zones translates into approximately +0.06 $mag_{SQM}/arcsec^2$ per year.

## 1 Introduction

Light pollution is a pervasive feature of the modern world. Assessing the long term evolution of the anthropogenic brightness of the night sky is a challenge that elicits great interest. Several light pollution monitoring networks have reported overall trends of darkening of their night skies, at rates about several hundredths of a magnitude per year [1–4], even in strongly light polluted urban locations where the yearly variations of airglow associated with the solar cycle are not expected to be a relevant factor. These trends can be due to multiple causes, including effective reductions of the artificial night sky brightness due to improved outdoor lighting installations and reductions of the lighting levels [5,6], spectral shifts of the source emissions modulated by ground reflectance and atmospheric propagation and scattering [7–9] or intrinsic changes in the detectors sensitivity, henceforth refereed as aging, either by progressive deterioration of the optics or/and of the electronics of the devices. Isolating these effects is crucial, in order to determine the actual trends of the artificial emissions. We report in this note some preliminary results on the aging of Sky Quality Meter (SQM) detectors permanently installed outdoors, obtained by means of a direct method, that is, by the comparison of their before/after readings with those of some reference detectors of the same kind that were not exposed to the environment in the intervening period of time. Our results strongly suggest that a relevant part of the darkening observed in our skies can be due to detector aging and should be compensated for. We have found a relatively well defined association between detector aging and total exposure to sunlight in several measurement networks that, in case of being confirmed, would allow to compensate with some easiness the present and historical SQM datasets and would help to refine the estimated underlying variation of the anthropogenic light emissions.

*S. Bará, E-mail address: salva.bara@usc.gal

## 2 Materials and methods

### *2.1 SQM sensors*

Five SQM units (Unihedron, CA) [10–12] belonging to different light pollution monitoring networks [13–16] and permanently deployed outdoors at fixed places were included in this study. All devices were installed with their Unihedron standard protective housing, with a soda-lime glass window. Their serial numbers and types, locations, periods of exposure to environmental conditions, and reference detectors used for drift assessment, are listed in Table 1.

### *2.2 Data acquisition*

The standard procedure to assess the detectors drift was to put a reference detector of the same kind continuously measuring in parallel, during a period of time from three weeks to three months, at the beginning and at the end of the elapsed period for which the aging of each detector was to be estimated. In the meantime, the reference detectors were either not used at all and kept isolated from the outdoor environment, or used in specific, short-duration measurement campaigns with exposure to nighttime radiance. Night sky brightness samples were taken at the characteristic rates of each network, ranging from one record every 40 s to one every five minutes, with a typical rate of one record per minute. This provided several thousands of useful samples per detector and measurement run that we refer to henceforth as a data segment.

Table 1. SQM units used in this study

| **Test Unit** | **Location** | **latitude / longitude** | **Assessment period** | **Reference Unit** | **Total drift** | **Total insolation** | **Drift rate** |
|---|---|---|---|---|---|---|---|
| serial #<br>and type | Name | WGS84<br>degrees | Beginning/End<br>yyyy-mm-dd | serial #<br>and type | $mag_{SQM}$/<br>$arcsec^2$ | MWh/<br>$m^2$ | $mag_{SQM}/arcsec^2$<br>per $MWh/m^2$ |
| 2750-LU | Otos | 38.8535<br>-0.4434 | 2014-10-18<br>2020-07-08 | 2708-LU | 0.39 | 10.137 | 0.039 |
| 2706-LU | València | 39.4781<br>-0.3761 | 2014-10-18<br>2020-09-24 | 2708-LU | 0.38 | 11.046 | 0.035 |
| 2587-LU | Tavernes de<br>la Valldigna | 39.0726<br>-0.2692 | 2014-10-18<br>2020-02-23 | 2708-LU | 0.31 | 9.666 | 0.032 |
| 2437-LE | Montsec | 42.0244<br>+0.7361 | 2016-05-10<br>2019-08-15 | 2749-LU | 0.22 | 5.836 | 0.038 |
| | | | 2016-05-10<br>2018-01-15 | 2749-LU | 0.08 | 2.916 | 0.026 |
| | | | 2018-01-15<br>2020-03-15 | 3797-LU | 0.11 | 3.627 | 0.030 |
| 2400-LR | Paramos | 43.0019<br>-8.6992 | 2014-06-19<br>2019-02-19 | LR Batch<br>average | 0.17 | 6.688 | 0.026 |

All LU units were fitted with Data Loggers (model SQM-LU-DL)

### *2.3 Data reduction*

The readings of the detector under test were linearly interpolated to the base of times of the samples acquired by the reference one. The samples corresponding to reference detector readings greater than or equal to 14.0 $mag_{SQM}/arcsec^2$ were selected for processing, independently from any other atmospheric or celestial factor, so that cloudy/cloudless and moonlit/moonless data were treated equally. The differences between the test and the reference readings were controlled in runtime to filter sporadic values greater than 1.0 $mag_{SQM}/arcsec^2$ that could be produced e.g. by moon shadow casting on one detector and not on the other. The average value and the standard deviation of both the sample and the mean of each data segment were subsequently calculated. These operations, carried out with the data obtained at the beginning of the exposure period, were performed again with the data acquired once this period had passed, and the differences after-before were evaluated as an estimate of the drift.

There was an exception to this standard procedure, in the Paramos detector of the Galician Night Sky Brightness Measurement Network. In that case we did not have a spare detector to be used as reference. Instead, the Paramos detector readings were compared with the average readings of 20 new SQM units installed together in 2014 for intercalibration, and with the average readings of another batch of 20 new SQM units installed at Paramos in 2019 for the same purposes. We assumed that the averages of both batches could be considered a reasonably reliable standard reference for detecting drifts in the detector under test.

The accumulated solar exposure was determined, for each measurement site and for the specific assessment period, on the basis of the monthly averages of global (i.e. direct plus scattered) solar irradiation over a horizontal plane, in $kWh/m^2$ per day, publicly available from the official CIEMAT website [17]. These values are calculated with a resolution of 5x5 km from satellite remote sensing data using standard models with ground-based validation from radiometers installed in a large network of meteorological stations (http://www.adrase.com/en/project.html).

## 3 Results

The overall change in sensitivity, the total exposure to global horizontal solar irradiation and the sensitivity drift per $MWh/m^2$ of exposure are listed in Table 1. The sensitivity drifts are plotted in Fig 1 against the solar exposure (left) and elapsed months (right). The slope of the linear fit is +0.036 (±0.002) $mag_{SQM}/arcsec^2$ per $MWh/m^2$. Expressed in terms of elapsed time, this amounts to +0.062 (±0.004) $mag_{SQM}/arcsec^2$ per year. The correlation seems to be slightly better defined versus exposure than versus elapsed time between measurements.

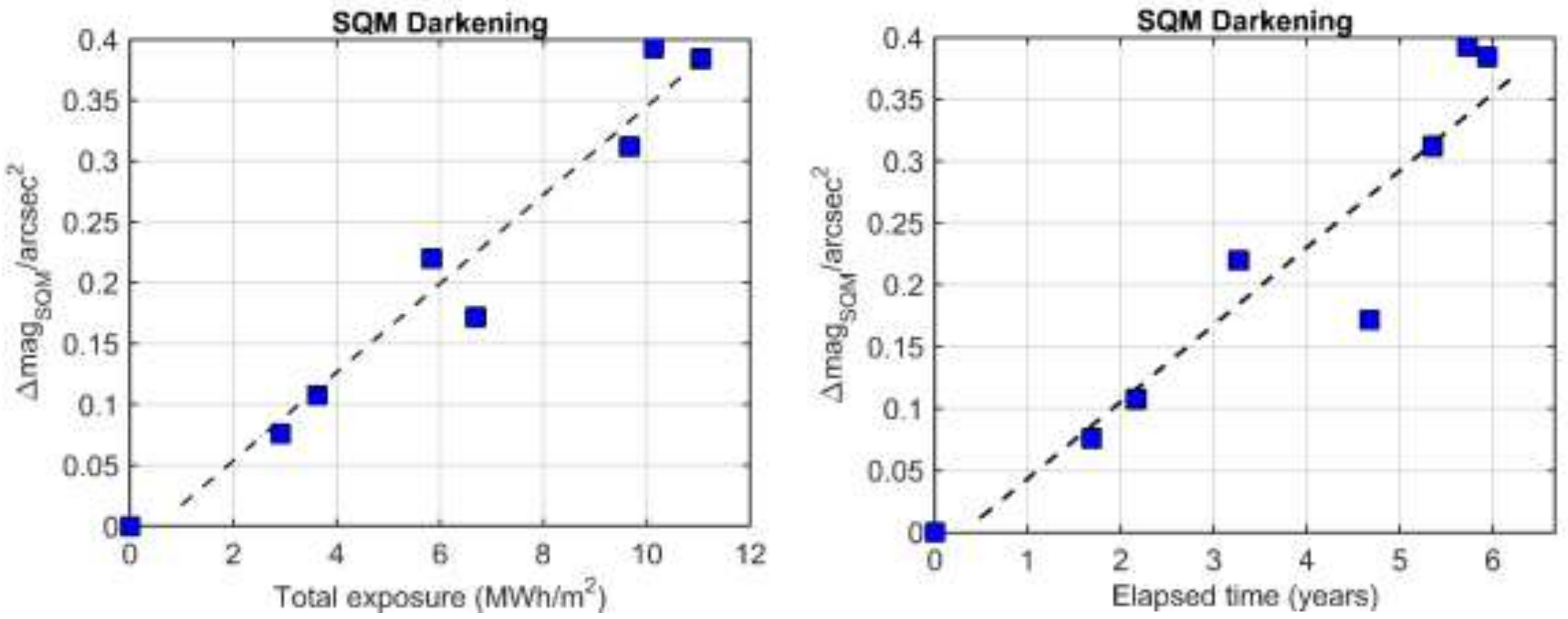

Fig. 1. Sensitivity drift (Table 1) plotted as a function of (left) global horizontal solar exposure and (right) elapsed time (months). The point at the origin reflects a necessary constraint of this problem

The A-type statistical uncertainty associated with the mean of the differences of the readings of each detector in each individual data segment is typically very small (or order ~0.001 $mag_{SQM}/arcsec^2$) due to the huge number of records used for determining that mean. B-Type uncertainty, associated with the nominal precision of the instrument, is also small. The often quoted value of 0.1 $mag_{SQM}/arcsec^2$ does not correspond to the uncertainty of the readings of a given instrument under constant illumination, but to the typical dispersion of the systematic differences among a large set of instruments. Since in our case we are studying the evolution of the differences of the readings of two fixed detectors, any initial systematic bias between them cancels out when determining the drift. We are left only with the instability of the readings of each unit under constant illumination, that we have found to be typically small, of order of the resolution of the instrument (0.01 $mag_{SQM}/arcsec^2$). Assuming a constant probability density function within this interval, this translates into an expected standard deviation 121/2 times smaller. The resulting combined uncertainty of the individual data segments, multiplied by a factor 21/2 to account for the fact that each determination of the drift is the difference of the means at the beginning and the end of the evaluation period, is still slightly smaller than the precision of the instrument. The total drifts in $mag_{SQM}/arcsec^2$ listed in column 6 of Table 1 are then given with two decimal places. Regarding solar exposure, the period of calculation was determined with a precision of 1 day, that for the measurement sites studied in this work corresponds to an average insolation of ~5 $kWh/m^2$.

## 4 Discussion

The purpose of this short note is to report the drifts in observed in a sample of SQM detectors permanently exposed to the outdoor environment, determined by means of a direct procedure consisting in measuring the evolution of the differences between their readings and the ones of reference detectors of the same kind which were kept reasonably free from environmental exposure in the intervening periods of time. We do not intend at this point to provide a comprehensive description of the optical and electronic components that may contribute to these drifts nor of the physical and photochemical processes that may give rise to this deterioration. Detailed studies on these issues are presently being carried out by several groups worldwide, including ours, and we expect they could be available soon for open dissemination and discussion in the light pollution measurement community. Our aim now is to recall the existence of this effect and to provide an estimate of its magnitude. Any potential drift is strongly relevant for the long-term assessment of the changes in the anthropogenic brightness of the sky. At the same time we would like to encourage all research groups having similar types of direct estimates to make public their results, in order to ascertain whether the trends observed in our detectors are representative of wider samples of units or should be refined or modified in some way.

The reported results correspond to a small sample of detectors, but were obtained independently by different observers using diverse detectors in several locations, in different dates, and during different periods of time. All data were obtained using the same kind of detectors, hence we avoid the additional errors that could be produced by shifts in the spectral power density of the night sky brightness if the reference detectors had not the same passband as the ones under test.

The data reduction procedure was kept intentionally simple since the problem under study seems to allow for it: systematic changes in sensitivity of identical detectors can be evaluated by simultaneous measurements of the brightness of the night sky under very different celestial and meteorological conditions [18], so no special selection of the input data was required (excepting the basic one of selecting the range of magnitudes of interest and controlling for sporadic events affecting only one of the detectors). This permitted processing the datasets in an automatic way with minimal intervention of the operators, and only after these results were calculated, the traces of the detector measurements were visually inspected to perform a posteriori quality check. No unexpected behavior was detected in any case and in consequence no data were reprocessed.

According to the results presented in section 3, the accumulated amount of solar exposure seems to describe the observed drifts slightly better than the mere elapsed time. If the driving variable is solar irradiation, the plot in function of MWh/m2 shall have the form of a unique straight line valid for all detectors, whereas the plot in function of time should show a set of straight lines passing through the origin, with different slopes depending on the average solar irradiation per year at each site. We have as of today strong evidence that this is the case, from data obtained using other kinds of reference detectors located at other latitudes (not shown here). Let us point out, however, that since not all our stations are located in places with exactly the same average insolation, this effect is also reflected in our figure. Whereas most of our detectors were located at highly insolated places in the Mediterranean area, with a very high and uniform insolation rate, the Galician station of Paramos, located in the stormy and cloudy Atlantic coast, was subjected to significantly less insolation per year than the remaining ones. This is the data point that appears significantly displaced toward the right, in the middle zone of the time-based plot, as expected from a point belonging to a straight line of smaller (time-)slope.

If Sun exposure were finally found to be the dominant cause of the drifts, the deterioration processes could be probably attributed to the plastic components of the unit (via photochemical processes in the UV and blue regions of the spectrum, and complementary thermal ones produced by the overall exposure). Of course, this should not be a matter of concern for those SQM detectors predominantly used at nighttime, that are not permanently exposed to sunlight. In case elapsed time itself were found to be more relevant, the observed trends could be related to deterioration of the electronics under continuous usage, dust deposition over the sensor window, and/or environmentally driven chemical processes in the optics not directly related to sunlight exposure. All these effects are of course not mutually exclusive and can be acting simultaneously on the detectors permanently deployed in the field.

Note finally that throughout this note we generically referred to the observed effect as a 'sensitivity drift'. This term adequately describes the change in the response of the device to the incident radiance integrated within the SQM spectral passband. When working with magnitudes, which are a logarithmic function of the in-band radiances, this effect could also be properly called a 'zero-point drift'.

## 5 Conclusions

Long-term monitoring of light pollution in large detector networks requires using calibrated equipment and the assurance that this calibration is kept within well-known and reliable limits along all data acquisition periods.

In this work we provide additional evidence on the long-term sensitivity drift of SQM sensors, an effect that can mask the true evolution of the artificial night sky brightness in light polluted places. This effect is added to the one produced by the spectral and intensity shift associated with the accelerated pace of replacement of traditional lighting systems by LED sources, that gives rise to different evolution rates if evaluated with detectors of different spectral passband. The results provided in this work were obtained using reference detectors of the same kind of the ones under test, avoiding that way formal bias due to passband differences.

The observed sensitivity shift seems correlate well with the accumulated global solar horizontal irradiation to which the detectors were exposed. The average drift is +0.036 (±0.002) $mag_{SQM}/arcsec^2$ per $MWh/m^2$. Expressed in terms of elapsed time, this amounts to +0.062 (±0.004) $mag_{SQM}/arcsec^2$ per year at the particular sites our stations were located. The data analyzed here were obtained at mid-latitudes under typical solar irradiation of ~5kWh/$m^2$ per day, averaged over the whole year. It would be particularly interesting analyzing the time-dependent behavior of data from places with significantly different values of solar exposure.

Sensitivity drifts may affect to all sensors permanently deployed outdoors with parts susceptible of weathering degradation, not only to the SQM. Periodic checks of the night sky brightness sensors with the use of reference detectors of the same type are highly advisable.

## Acknowledgments

This work was supported in part by Xunta de Galicia, grant ED431B 2020/29. J.Z. acknowledges the support from ACTION, a project funded by the European Union H2020-SwafS-2018-1-824603, RTI2018-096188-B-I00 and S2018/NMT-4291 (TEC2SPACE-CM). Part of this work was developed in the framework of the Spanish Network for Light Pollution Studies (REECL).